# Multidimensional Fairness in Paper Recommendation

Reem Alsaffar[1], Susan Gauch[1] and Hiba Al-Kawaz[2]

[1]University of Arkansas, Fayetteville, AR, USA
[2]Baghdad University, Baghdad, Iraq
[1]{rbalsaff, sgauch}@uark.edu
[2]hiba.m@csw.uobaghdad.edu.iq

**Abstract**: To prevent potential bias in the paper review and selection process for conferences and journals, most include double blind review. Despite this, studies show that bias still exists. Recommendation algorithms for paper review also may have implicit bias. We offer three fair methods that specifically take into account author diversity in paper recommendation to address this. Our methods provide fair outcomes across many protected variables concurrently, in contrast to typical fair algorithms that only use one protected variable. Five demographic characteristics—gender, ethnicity, career stage, university rank, and geolocation—are included in our multidimensional author profiles. The Overall Diversity approach uses a score for overall diversity to rank publications. The Round Robin Diversity technique chooses papers from authors who are members of each protected group in turn, whereas the Multifaceted Diversity method chooses papers that initially fill the demographic feature with the highest importance. We compare the effectiveness of author diversity profiles based on Boolean and continuous-valued features. By selecting papers from a pool of SIGCHI 2017, DIS 2017, and IUI 2017 papers, we recommend papers for SIGCHI 2017 and evaluate these algorithms using the user profiles. We contrast the papers that were recommended with those that were selected by the conference. We find that utilizing profiles with either Boolean or continuous feature values, all three techniques boost diversity while just slightly decreasing utility or not decreasing. By choosing authors who are 42.50% more diverse and with a 2.45% boost in utility, our best technique, Multifaceted Diversity, suggests a set of papers that match demographic parity. The selection of grant proposals, conference papers, journal articles, and other academic duties might all use this strategy.

## 1 Introduction

Our modern world is very diverse and individuals and institutions strive for inclusion. Despite gain in legislation and attitudes, there is still discrimination against people because of their race, color, gender, religion, national origin, disability or and age [1]. These groups have legal protection within the United States, but discrimination, conscious or unconscious, still exists throughout society [38, 13]. Unfortunately, academia is no exception as evidenced by the fact that only 38% of tenure-track positions were awarded to women despite women receiving more than 50% of the Ph.D.'s awarded [15]. Computer Science is a long way from achieving diversity. [45] and [10] document the fact that only 18% of graduates in Computer Science are women and also only 18% are minorities. These statistics are reflected in the lack of diverse speakers at Computer Science conferences and in the demographics of conference attendees where minorities are underrepresented [28]. Racial, gender, and other types of discrimination among reviewers, editors and program committee might lead to bias in choosing papers for publishing [33] which has led to SIGCHI, one of the highest impact ACM conferences, announcing in 2020 an explicit goal of increasing the diversity of its Program Committee [40]. Merely using a double-blind review process fails to solve the problem of discrimination [9, 30]. Reviewers can often infer the authors of papers from previous publications or readily available electronic preprints even when using double-blind review, so the review process is not actually double-blind [1, 36]. Our approach is based on building a profile for each paper that reflects the paper's overall quality and also models the diversity of the paper authors. The demographic features most frequently identified as a source of bias are Gender, Ethnicity [7], Career Stage [31], University Rank [16] and geolocation [26], thus these are the features we use in our demographic user profiles. Our fair recommender system then uses this multi-dimensional profile to recommend papers for inclusion in the conference balancing the goals of increasing the diversity of the authors whose work is selected for presentation while minimizing any decrease in the quality of papers presented.

In our previous paper, we proposed two methods that use multiple attributes when picking a subset of authors from the pool to achieve demographic parity. We incorporated the diversity and the quality of the authors during the selection process to minimize the utility loss and maximize the diversity. We applied two different feature weighting schemes, Boolean and continuous, in demographic profiles used to increase fairness in paper recommendation. In this paper, we present three fair recommendation algorithms that balance two aspects of a paper, its quality and the authors' demographic features, when recommending papers to be selected by the conference. Because information about the review process is generally confidential, we simulate the results of the review process by creating pools of papers from related conferences within a specific field that have different impact factors. The highest impact factor conference papers will play the role of the papers that are rated most highly by the reviewers, the middle impact factor conference papers those with the second best reviews, and papers published at the conference with the lowest of the three impact factors will be treated as papers with lower reviews. Our main contributions in this work are:

- Modelling author demographics using profiles that contain multiple demographic features.
- Developing and evaluating fair recommendation algorithms for paper selections that balance quality and diversity.
- Achieving demographic parity between the accepted authors with the pool of all authors.

## 2   Related Work

We begin by discussing aspects of bias in academia, then we review previous work on the construction of demographic profiles for users. Finally, we summarize recent approaches to incorporate fairness in algorithmic processes.

### 2.1   Bias

**Bias in Academia:** Bias in academic research can be seen when one outcome or result is preferred over others during the testing or sampling phase, and also during any research stage, i.e., design, data collection, analysis, testing and publication [37]. Bornmann and Daniel discuss the evidence that gender, major field of study, and institutional affiliation caused bias in the committee decisions when awarding doctoral and post-doctoral research fellowships [5]. Flaherty investigates discrimination in the US college faculty focusing on ethnicity. The results showed that only 6% of professors are black versus 76% of white professors [17]. More recently, an article published by researchers from Stanford Graduate School of Education in 2021 showed that, in the United States, more doctoral degrees have been earned by women than men. Despite this, women are still less likely than men to receive tenured positions, have their research published, or obtain leadership roles in academia. After analyzing one million doctoral dissertations from US universities, they found that the authors whose topics are related to women or who used methodologies that refer to women have decreased career prospects versus those related to men [51].

**Bias in Peer Review:** Several studies have that the lack of fairness in the peer review process has a major impact on which papers are accepted to conferences and journals [41]. Reviewers tend to accept the papers whose authors have the same gender and are from the same region [33]. Double-blind reviews do not entirely solve this issue and some researchers demonstrate that bias still exists in the reviewing process. For example, Cox et al. concludes that the double-blind review did not increase the proportion of females significantly compared with the single-blind review [10].

## 2.2 Fairness

**Demographic Profiling:** User profiling can be used to understand the users' intentions and develop personalized services to better assist users [18]. Recently, researchers are incorporating demographic user profiles in recommender systems hoping to limit unfairness and discrimination within the recommendation process [29, 14]. Within academia, the demographic attributes of age, gender, race, and education are widely used and researchers often infer these features from the user's name [8, 39].

**Demographic Parity:** To achieve fairness, many approaches aim for *demographic parity*, which is when members of the protected groups and non-protected groups are equally likely to receive positive outcomes. However, this requirement generally causes a decrease in utility. Yang and Stoyanovich focus on developing new metrics to measure the lack of demographic parity in ranked outputs [46]. [47] and [48] address the problem of improving fairness in the ranking problem over a single binary type attribute when selecting a subset of candidates from a large pool while we work with multiple features at the same time. It maximizes utility subject to a group fairness criteria and ensuring demographic parity at the same time. We extended these works by using multiple attributes when picking a subset of authors from the pool to achieve demographic parity. We also incorporated the diversity and the quality of the authors during the selection process to minimize the utility loss and maximize the diversity. A study by [52] extended these works by using multiple attributes when picking a subset of authors from the pool to achieve demographic parity. Using Boolean profiles, they also incorporated the diversity and the quality of the authors during the selection process to minimize the utility loss and maximize the diversity. Further extensions [53] compared two different feature weighting schemes, Boolean and continuous, in demographic profiles used to increase fairness in paper recommendation [52]. Recently, some authors have been working on Generalized Demographic Parity (GDP) which is a group fairness metric for continuous and discrete features, to make fairness metrics more accurate. [54] proposed their method by displaying the relationship between joint and product margin distributions distance. They demonstrate two methods named histogram and kernel with linear computation complexity. Their experiment showed that GDP regularizer can reduce bias more accurately.

**Fairness in Machine Learning:** As more and more algorithms make financial, scholastic, and career decisions, it is very important that the algorithms to not perpetuate bias towards demographic subgroups. Many investigations show that machine learning approaches can lead to biased decisions [12]. Thus, researchers are working to improve classifiers so they can achieve good utility in classification for some purpose while decreasing discrimination that can happen against the protected groups by designing a predictor with providing suitable data representation [22, 49]. Other researchers attempt to improve fairness by training machine learning models without knowing the protected group memberships. In particular, [55] proposed an Adversarially Reweighted Learning (ARL) approach to improve the utility for the least represented protected groups when they train the model. During the training stage, they relied more on the non-protected features and task labels to identify unfair biases and train their model to improve fairness. Their solution outperformed state-of-the-art alternatives across a variety of datasets.

**Paper Assignment Fairness:** Some researchers have explored and measured fairness when choosing a suitable reviewer to review a paper. [32] and [42] focus on fairness and statistical accuracy in assigning papers to reviewers in conferences during the peer review process. Most of these studies propose methods to improve the quality of the reviewer assignment process. We contribute to this area by creating author profiles with multiple demographic features and using them in new fair recommendation algorithms to achieve demographic parity when selecting papers for inclusion in a conference.

## 3 Demographic Profile Construction

We first build a demographic profile for each paper by modeling the demographic features for the paper's authors so that this information is available during paper selection. Some demographic features are protected attributes, e.g., gender, race, that qualify for special protection from discrimination by law [25]. In this section, we will describe how we collect the demographic features for each author in our papers pool and then how we build the paper profile.

### 3.1 Data Extraction

For a given paper, our goal is to extract five demographic features that are Gender, Race, University Rank, Career Stage, and Geolocation for its author(s) then combine them to create a profile for the paper. Each feature is mapped to a Boolean value, either 1 (true) or 0 (false) based on that paper's author(s) membership in the protected group. We then extend our approach beyond current approaches by modeling demographics with continuous-valued features (each feature is mapped to a value between 0 and 1) to represent the complement of the proportion of each feature among computer science professionals. Table 1 outlines the protected and non-protected categories for each of our demographic features.

**Table 1.** Demographic Features and Categories

| Feature | Protected/Non-Protected Category |
|---|---|
| Gender | Female / Male |
| Ethnicity | Non-White / White |
| Geo-Location | Developing /Developed (by country) |
| | EPSCoR / Non-EPSCoR (by state in USA) |
| Career Stage | Junior / Senior |
| UniversityRank | Less than or equal mean/ more than mean |

**Gender**: To gather information about an author's gender, we use the NamSor API v2, a data mining tool that uses a person's first and last names from different languages, alphabets, countries, and regions to infer their gender. The software processed more than 4 billion names with high precision and recall which are 98.41% and 99.28% respectively. The tool returns a value between -1 and +1 indicates that the name is male if it is close to +1 and female if it is close to -1. The accuracy of gender prediction using this tool is close to 99% [3]. After collecting each author's gender, we map females to 1 since they are the protected group and males to 0. To calculate the continuous value for gender, we map females and males to the complement of their participation in computer science. Women are considered a protected group since they make up only 27% of professionals in the computer science field [4]. **Ethnicity**: To predict ethnicity, we again use the NamSor tool, a web API that is used to predict ethnicity from the first and last names with the limitation of 500 names/month. It returns ethnicity as one of five values: {White, Black, Hispanic, Asian, other} [3]. Non-whites are considered a protected group since they make up less than 40% of professionals in the computer science field [4]. The continuous values for Ethnicity were calculated by mapping each category to the complement of its proportion in the population of Computer Science professionals from [51] (Computer, engineering, & science occupations, 2020). Whites comprise 70.46% of computer science professionals, so they are mapped to 0.2954. Similarly, Black, Asian, Hispanic and others are assigned to 0.9295, 0.8237, 0.9281, and 0.7400 respectively. **Career Stage:** In order to extract the academic position for each author, we utilize the researcher's Google Scholar pages [19] or their homepages. Researchers whose primary appointment is within industry are omitted from our data set. The results are then mapped to Boolean values, 0 if they are a senior researcher (Distinguished Professor, Professor, Associate Professor) and 1 if they are a junior researcher (Assistant Professor, Postdoc, Student). To calculate the continuous values for this feature, we map to six values equally distributed between [0, .., 1.0] in increasing order by rank, i.e., Distinguished Professor: 0/5 = 0.0; Professor: 1/5 = 0.2; ...; Student: 5/5 = 1.0. Table 2 shows the values for each category.

**Table 2.** Career Stage Weight Allocation

| Position | Weight |
|---|---|
| Distinguished Professor | 0.17 |
| Professor | 0.33 |
| Associated Professor | 0.50 |
| Assistant Professor or Lecturer | 0.67 |
| Post-Doctoral or Research Fellow | 0.83 |
| Graduate Student | 1.0 |

**University Rank:** Collecting this feature is done by extracting the institution's name from the Google Scholar page for the author [19] or their home pages and mapping it to the World University Rankings obtained from [44]. We partition the authors into low-rank (1) or high-rank institutions (0) using the median value. Then, we normalize the raw value to a continuous value by dividing the university rank ($U_r$) by the lowest university rank ($L_r$):

$$R_C = \frac{U_r}{L_r} \quad (1)$$

**Geolocation:** We set the researcher's geolocation (country and state if inside the US) based on information extracted from their institution's home page using the university name that was extracted. If the author is working inside the United States, we extract the state name as well. We find the category of the country (developed or developing) by mapping the country to the tables of the developed and developing economies that offered by the UN [35]. Thus, the Geolocation Boolean value is assigned to 0 if the researcher is working in a developed country (non-protected group) and 1 if a developing country (protected group). For those who live in the US, we use the EPSCOR (Established Program to Simulate Competitive Research) [34] to map the Geolocation to Boolean values. EPSCoR states which obtain less federal grant funding are the protected group with the value 1 and non-EPSCoR states values are 0. To calculate the continuous value for the Geolocation, we use the complement values of Human Development Index (HDI) ranking [24]. The values are ranging from 0.957 to 0.394 and Table 3 shows a sample of these values.

**Table 3.** Countries HDI sample

| Country | HDI |
|---|---|
| Norway | 0.957 |
| Ireland | 0.955 |
| Iceland | 0.949 |
| Germany | 0.947 |
| Sweden | 0.945 |
| Australia | 0.944 |

**H-index:** We extract the h-index for each author from their Google Scholar page so we can measure the conference utility in our evaluation. If the author doesn't have a scholar page, we obtain their h-index using Harzing's Publish or Perish tool. This software calculates the h-index for the scholar using some impact metrics [23].

To conclude, each researcher has a demographic profile consists of five features (gender, ethnicity, career stage, university rank, and geolocation). Each feature has a Boolean weight that represents whether or not the candidate is a member of the protected group for that feature and a continuous value to represent the complement of the proportion of each feature among computer science professionals. In addition, we collect the h-index for each researcher using either their Google Scholar profile or is calculated and we use it to evaluate the utility of each accepted papers list in our evaluation.

### 3.2 Paper Profile Formation

We construct the demographic profile for each paper by combining the demographic profiles for all of the paper authors. Recall that each author has a Boolean value profile and a continuous value profile.

**Boolean:** The paper profile is created by doing a bit-wise OR on the paper's author profiles. Thus, the paper profile is 1 for a given demographic feature when any author is a member of that feature's protected group. We considered summing the author profiles, but this would give preferential treatment to papers with more authors and normalizing the summed profile would penalize papers with many authors.

**Continuous:** The paper's demographic profile is created by selecting the maximum value for each feature among the paper authors' profiles.

### 3.3 Paper Quality Profiler

There are several ways to measure a paper's quality such as the number of citations of the paper, the reputation of the editorial committee for the publication venue, or the publication venue's quality itself, often measured by Impact Factor (IF) [6]. Although the IF is not accurate for new venues that contain high quality papers with few citations, we use it as the basis of the quality profile for the papers in our research since the conferences in our dataset are all well-established [50]. We extract the Impact Factor (IF) for each paper's conference from Guide2Research website published in 2019 [21]. The IF was calculated by using Google Scholar Metrics to find the highest h-index for the published papers in the last 5 years [20].

### 3.4 Pool Distribution

When applying our proposed methods as described below, we rely on reaching demographic parity during accomplishing our goal. This means that we select the papers such that the demographics of the accepted authors match those of the pool of candidates. To achieve this, we measure the proportion of participants for each feature in the pool and store them in a vector (PoolParity).

$$PoolParity=<GenderWt, EthnicityWt, CareerWt, UniversityWt, GeoWt>$$

where each weight is the number of authors from that protected group normalized by the number of authors in the pool.

## 4 Approaches

The next goal is maximizing the diversity of the conference by applying three different methods to select papers with respect to each features' distribution in the pool and achieving demographic parity. The reason is to get a list of papers that have more diverse people in the high rank conferences while keeping the level of quality the same or with a little drop.

### 4.1 Overall Diversity Method

After creating paper demographic profiles as described in section (3), paper diversity scores (PDScore) are calculated using formula (2) on the feature values:

$$PDScore = \sum_{i=1}^{5} f_i \qquad (2)$$

where $f_i$ is the value for each paper's demographic feature (i.e., five features for each paper). Our first method to choose a diverse list of papers considers two different queues. The quality queue ($Qquality$) which contains the papers ranked by the Impact Factor (IF) as described in Section 3. This gives preference to the papers ranked highest by the reviewers, in our case represented by papers that appeared in the most selective conference. The demographic queue ($Qdemog$) which contains the ranked papers by PDScore. Next, we pick papers from the top of ($Qdemog$) until satisfying the pool demographic parity for each feature then the remaining papers are added from the quality queue in order to meet the number of papers desired by the conference. Thus, as long as there are sufficient candidates in the pool, we are guaranteed to meet or exceed demographic parity for each protected group.

**Algorithm 1:** Overall Diversity

1   $Qq$uality, $Qd$emog ← Initialize two empty priority queues
2   PoolParity ← Initialize an empty vector
3   $Qq$ ← insert the papers and sort them based on Quality-Scores
4   for each feature:
5      PoolParity [feature] ← compute Demographic Parity
6   for each paper:
7      PDScore ← compute paper diversity score
8      add paper to $Qd$emog and order them using PDScore
9      If 2 or more papers have same PDScore:
10        Sort papers using Quality-Score
11 while PoolParity Not satisfied:
12    Papers ← select a paper from top of $Qd$emog
13    delete selected paper from $Qq$uality
14 while # of conference papers not satisfied:
15    Papers ← select a paper from top of $Qq$uality

## 4.2 Multi-Faceted Diversity Method

The previous method selects papers based on the total diversity score for each paper. However, it does not guarantee that the selected authors from the protected groups are actually diverse. It might end up selecting papers that have high diversity scores but are all females from developing countries, for example, with no minority authors at all. To correct for this possibility, we extend the previous approach by creating five ranked queues (one per feature) and sorting the papers using one demographic feature at a time. Including the quality-ranked queue, we now have six queues total. Based on the pool demographics, we give the highest priority to the rarest features in the pool first, so we create the accepted papers list by selecting papers from the queues whose features have the fewest candidates in the pool until the demographic parity goal for those features is achieved. After satisfying demographic parity for all protected groups, the remaining papers are added in order from the quality queue.

**Algorithm 2:** Multi-Faceted Diversity

1   FeatureName ← List of five queue names, one per feature
2   for each feature in FeatureName:
3      DivQueue[feature] ← Initialize empty priority queue
4   $Q$ualityQueue ← Initialize an empty priority queue
5   PoolParity ← Initialize an empty vector
6   $Q$ualityQueue ← insert papers and sort by Quality-Score
7   for each feature in FeatureName:

8    PoolParity [feature] ← compute Demographic Parity
9  for each paper:
10    PDScore ← compute paper diversity score
11 for each feature in FeatureName:
12    DivQueue[feature] ← add paper if this feature is 1
13    Sort papers based on Quality-Score
14    If 2 or more papers has the same Quality-Score:
15        Sort papers using PDScore
16  while PoolParity NOT empty:
17    LowFeature ← min (PoolParity)
18    while LowFeature Not reached demographic parity
19       Papers ← select top DivQueue[LowFeature]
20       delete selected paper from $Q$ualityQueue
21  delete LowFeature from DParity
22  while # of conference papers not satisfied:
23    Papers ← select a paper from top of $Q$ualityQueue

### 4.3 Round Robin Method

Our third method employs the same demographic queues as the Multi-Faceted Diversity method, but visits the queues using a round robin algorithm. It specifically creates one priority queue per demographic feature and selects from each in a round-robin fashion until the list of papers is formed. Since the proposed demographic profiles model five features, we create five priority queues, each of which orders all the papers by one specific feature based on PDScore. For example, one queue orders the papers based on gender, whereas another would order the papers based on academic position. Once all queues have been sorted, we apply round-robin selection by picking the highest sorted profile from each queue. Accordingly, the selected profile is then eliminated from all queues to endure that the same profile will not be selected again. This process continues iteratively until the list of papers reaches the desired size. We use this approach with both Boolean and continuous weight diversity profiles.

**Algorithm 3:** Round Robin

1 FeatureName ← List of five queue names, one per feature
2 for each feature in FeatureName:
3       DivQueue[feature] ← Initialize empty priority queue
4 $Q$ualityQueue ← Initialize an empty priority queue
5 $Q$ualityQueue ← insert papers and sort by Quality-Score
6 for each paper:
7       PDScore ← compute paper diversity score
8 for each feature in FeatureName:

```
9        add profile to [feature] using PDScore[feature] as priority order
10  Papers ← empty list
11  While number of Papers < N:
12        feature ← FeatureName[0]
13        repeat:
14            paper ← get and remove profile from DivQueue[feature]
15        until paper  is not in Papers
16        add paper  to Papers
17        delete feature from FeatureName.
18  we now have N papers selected.
```

## 5    Experiment and Result

We now introduce our dataset and describe the process of evaluating our algorithms.

### 5.1    Datasets

For our driving problem, we focus on selecting papers for a high impact computer science conference from a pool of papers that vary in quality and demographics. To create pools of candidate papers that simulate the papers submitted to a conference, we select a trio of conferences based on several criteria: 1) the conferences should publish papers on related topics; 2) the conferences should have varying levels of impact {very high, high and medium} mimicking submitted papers reviewed as *high accept, accept, borderline accept*; 3) the conferences should have a reasonably large number of accepted papers and authors. Based on these criteria, we selected SIGCHI (The ACM Conference on Human Factors in Computing Systems), DIS (The ACM conference on Designing Interactive Systems), and IUI (The ACM Conference where the Human-Computer Interaction (HCI) community meets the Artificial Intelligence community). The papers published in SIGCHI represent papers rated highly acceptable by SIGCHI reviewers, DIS papers represent papers rated acceptable by SIGCHI reviewers, and IUI papers represent papers rated borderline acceptable. Excluding authors from industry, we create a dataset for each conference that contains the accepted papers and their authors (see Table 4). This dataset contains 592 papers with 813 authors for which we demographic profiles.  We will expand this work to other conferences in the future.

**Table 4.** Composition of Our Dataset [56].

| Dataset | Accepted Papers | Authors | Impact Factor |
|---|---|---|---|
| SIGCHI17 | 351 | 435 | 87 |
| DIS17 | 114 | 231 | 33 |
| IUI17 | 64 | 147 | 27 |

The demographic distribution of the authors in each conference is summarized in Fig. 1. These clearly illustrate each of the conferences had few authors from most of the protected groups with the lowest participation in the highest impact conference, SIGCHI, with gender being an exception. As an example, SIGCHI 2017 had only 8.28% non-white authors, DIS 2017's authors were only 16.45% non-white, and IUI 2017 had 27.21% non-white.

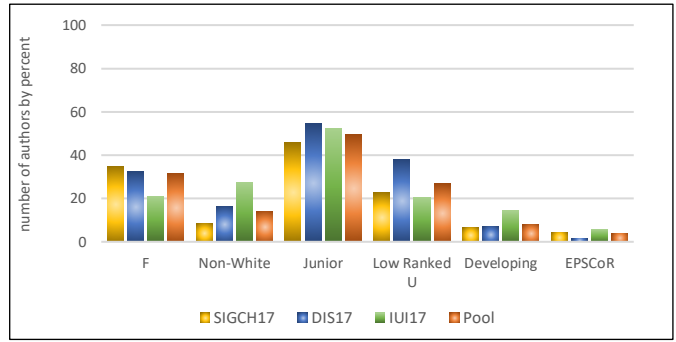

**Fig. 1.** Protected Group Membership of Authors for Three Current Conferences [56].

We define demographic parity as the participation rate for each of our demographic features in the pool created by combining the authors of all three conferences. Based on the 813 authors in our dataset, Table 5 presents the average participation in the pool for each feature and thus the demographic parity that is our goal.

**Table 5.** Demographic Participation from protected groups in Three Current Conferences [56].

|         | Gender  | Ethnicity | CStage  | URank   | Geoloc  |
|---------|---------|-----------|---------|---------|---------|
| SIGCHI  | 45.01%  | 7.69%     | 52.14%  | 25.64%  | 8.26%   |
| DIS     | 57.89%  | 31.58%    | 72.81%  | 55.26%  | 11.40%  |
| IUI     | 39.06%  | 56.25%    | 76.56%  | 28.13%  | 26.56%  |
| Average | 47.07%  | 18.71%    | 59.55%  | 32.33%  | 11.15%  |

### 5.2 Baseline and Metrics

*Baseline.* Our baseline is the original list of papers that were chosen by the program committee for SIGCHI 2017 and were represented in the venue. As shown in Table 2, the distribution of the protected groups in our baseline is: 45.01% female, 7.69% non-white, 52.14% junior professors, 25.64% authors from low ranked universities and 8.26 authors from developing countries.

*Metrics.* We evaluate our algorithms' effectiveness by calculating Diversity Gain ($D_G$) of our proposed set of papers versus the baseline:

$$D_G = \frac{\sum_{i=1}^{n} MIN(100, \rho_{G_i})}{n} \qquad (3)$$

where $\rho_{G_i}$ is the relative percentage gain for each feature versus the baseline, divided by the total number of features $n$. Each feature's diversity gain is capped at a maximum value of 100 to prevent a large gain in a single feature dominating the value.

By choosing to maximize diversity, it is likely that the quality of the resulting papers will be slightly lower. To measure this drop in quality, we use the average h-index of the paper authors and compute the utility loss ($UL_i$) for each proposed list of papers using the following formula:

$$UL_i = \frac{U_b - U_{P_j}}{U_b} * 100 \qquad (4)$$

where $U_{P_i}$ is the utility of the proposed papers for conference i and $U_b$ is the utility of the baseline. We then compute the utility savings ($Y_i$) of papers for conference i relative to the baseline as follows:

$$Y_i = 100 - UL_i \qquad (5)$$

We compute the F measure [27] to examine the ability of our algorithms to balance diversity gain and utility savings:

$$F = 2 * \frac{D_G * Y_i}{D_G + Y_i} \qquad (6)$$

In order to measure how far away from demographic parity our results are, we calculate the Euclidean Distance [11] between our selected papers and the pool:

$$\text{DemographicDistance} = \sqrt{\sum_{i=1}^{5}(F1_i - F2_i)^2} \qquad (7)$$

where F1 is the participation of each feature in the proposed list of papers to select and F2 is the feature's participation in the pool. Finally, we normalized the distance values to obtain the similarity percentages between our results and the pool as shown in the formula below:

$$\text{DemographicSimilarity} = 1 - \frac{\text{DemographicDistance}}{\text{MaxD}} \qquad (8)$$

where MaxD is the largest possible distance between two vectors in our feature space.

To summarize the ability of the methods to balance the competing demands of increasing demographic parity and saving utility, we again apply the F measure using formula 6 calculated using DemographicSimilarity and $Y_i$.

### 5.3 Results

Our recommender system produces ranked list(s) from which we select to form the accepted papers list with the overarching goal of increasing the diversity in the papers. Both methods reported here select papers from a quality sorted queue and one or more demographic queue(s). Whenever there are ties in a demographic queue, those papers are sorted by their quality score.

## 5.4 Comparison with the Baseline

We report the differences between the accepted papers in SIGCHI 2017 and the accepted papers produced by the recommender system described in Section 4 using Boolean and Continuous profiles. Looking at Fig. 2., we can see that all algorithms succeeded in increasing the diversity in the recommended papers for acceptance across all demographic groups when using the Boolean profiles. However, it is obvious that Round Robin method produced the highest diversity in all the protected groups except gender.

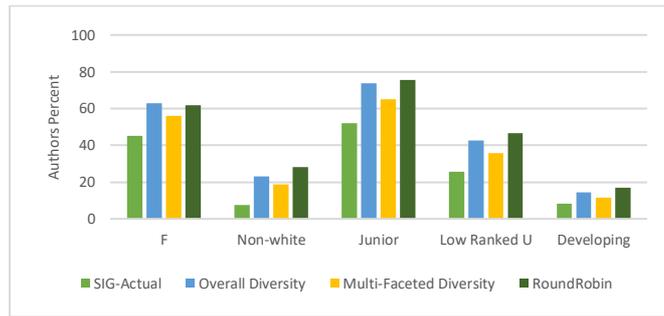

**Fig. 2.** Improvement in Protected Group Participation between the SIGCHI2017 and our Paper Recommendation Algorithms when using Boolean Profiles.

Fig.3 represents the protected groups participation with the Continuous profiles when applying our proposed recommendation algorithms. We can see that all methods succeeded in increasing the diversity in the recommended papers for acceptance across all demographic groups.

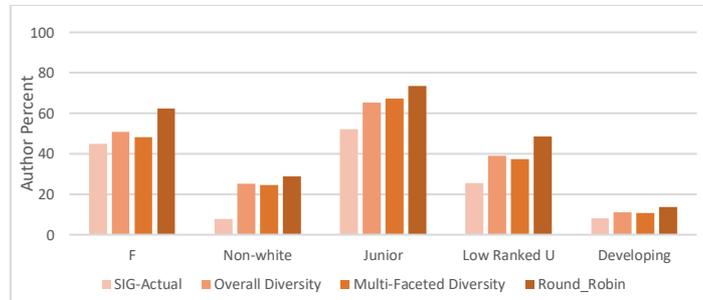

**Fig. 3.** Improvement in Protected Group Participation between the SIGCHI2017 and our Paper Recommendation Algorithms when using Continuous Profiles.

Table 6 compares the participation of the protected groups between the actual accepted papers for SIGCHI with the accepted papers proposed by our three methods, and demographic parity based on the participation of the protected groups in the pool of authors in our dataset. We can see that all algorithms increase the diversity of authors across all protected groups. With the exception of female researchers for the Boolean profile, the Round Robin algorithm increases participation among the protected groups more than the Multifaceted Diversity algorithm across all demographics. As expected, these diversity-based recommendation methods overcorrected by including more authors from the protected groups proportionally than in the pool as a whole.

Table 6. Protected Group Participation for the recommender algorithms using Boolean and Continuous profiles.

| Feature | SIGCHI | Overall Diversity (B) | Overall Diversity (C) | Multi-Faceted (B) | Multi-Faceted (C) | Round Robin (B) | Round Robin (C) | Pool |
|---|---|---|---|---|---|---|---|---|
| Female | 45.01% | 62.96% | 50.71% | 56.13% | 48.15% | 61.82% | 62.39% | 47.07% |
| Non-White | 7.69% | 23.08% | 25.36% | 18.80% | 24.50% | 28.21% | 28.77% | 18.71% |
| Junior | 52.14% | 73.79% | 65.24% | 64.96% | 67.24% | 75.50% | 73.50% | 59.55% |
| Low Ranked University | 25.64% | 42.45% | 39.03% | 35.90% | 37.32% | 46.44% | 48.43% | 32.33% |
| Develop Country | 8.26% | 14.53% | 11.11% | 11.68% | 10.83% | 16.81% | 13.68% | 11.15% |

The recommended papers are a mix of papers from the three conferences in our datasets in different proportions as described in Table 7. The Multi-Faceted Diversity method selects the highest proportion of the recommended papers, 85.8% (Bool) and 78.06% (Cont.), from the actual SIGCHI papers, but Overall Diversity and Round Robin also selects the majority of its papers, 75.5% (Bool), 62.11% (Cont.) and 61.53% (Bool), 60.11% (Cont.) respectively, from the original SIGCHI selected papers. We further observe that Overall Diversity and Multi-Faceted algorithms selected the majority of papers from the demographic queue(s) with only a few from the quality-sorted queue. The Overall Diversity method selected 67.24% (Bool) and 66.67% (Cont.) of its accepted papers from the demographic queue and only 32.76% (Bool) and 33.33% (Cont.) from the quality queue. In contrast, the Multi-Faceted Diversity method selected nearly all of its accepted papers, 92.88%, from one of the five demographic queues, and only 7.12% from the quality queue.

Table 7. Proportion of Recommended Papers from each Conference.

|  | Overall Diversity (B) | Overall Diversity (C) | Multi-Faceted (B) | Multi-Faceted (C) | Round Robin (B) | Round Robin (C) |
|---|---|---|---|---|---|---|
| SIGCHI | 265 (75.5%) | 218 (62.11%) | 301 (85.8%) | 274 (78.06%) | 216 (61.53%) | 211 (60.11%) |
| DIS | 59 (16.8%) | 87 (24.79%) | 47 (13.4%) | 61 (17.38%) | 84 (23.93%) | 90 (25.64%) |
| IUI | 27 (7.7%) | 46 (13.11%) | 3 (0.9%) | 16 (4.56%) | 51 (14.52%) | 50 (14.25 %) |
| Papers # | 351 | 351 | 351 | 351 | 351 | 351 |

We also compare the performance of our algorithms with respect to the quality of the resulting accepted papers. Table 8 summarizes the diversity gain ($D_G$), Utility Savings ($Y_i$), and F scores for the accepted papers proposed by each algorithm when using the Boolean and Continuous profiles. All methods obtained Diversity Gains of over 40% for the proposed set of accepted papers, with the biggest gain occurring with the Round

Robin algorithm. The gains in diversity occur with Utility Savings of 95.14% (B) and 94.06 (C) for the Round Robin algorithm versus 93.47% (B), 97.52%(B) and 97.52% (C), 102.45% (C) for the Overall Diversity and Multi-Faceted Diversity algorithms respectively. Based on these results, we conclude that the Round Robbin algorithm outperforms the other two algorithms and when considering author demographics and aiming for demographic parity, the quality of the selected papers actually increased.

**Table 8.** diversity gain and utility savings for our algorithms versus the Baseline for Boolean and Continuous profiles.

|  | Overall Diversity | Multi-Faceted Diversity | Round Robin |
|---|---|---|---|
| $D_G$ (Bool) | 64.58% | 46.00% | 72.65% |
| $Y_i$ (Bool) | 93.47% | 97.52% | 95.14% |
| F-score (Bool) | 76.39 | 62.51 | 82.39 |
| $D_G$ (Cont.) | 44.90% | 42.50% | 66.80% |
| $Y_i$ (Cont.) | 102.49% | 102.45% | 94.06% |
| F-score (Cont) | 62.44 | 60.08 | 78.12 |

Diversity-based algorithms may overcorrect and result in reverse discrimination, or the diversity gains may all be in one subgroup while other underrepresented populations are ignored. Tables 9 and 10 show the results when evaluating our algorithms' ability to achieve demographic parity with Boolean and Continuous features, respectively. We observe that, based on this criteria, the Multi-faceted Diversity algorithm produces results closest to Demographic Parity, with 95.01% similarity to the pool and a utility loss of just 2.48% when using Boolean profiles.

We further observe that the Multi-faceted method produces even better Demographic Parity of 95.12% when using continuous-valued features and actually results in a 2.45% increase in utility. This means that, by considering author diversity and aiming for demographic parity when selecting papers, the quality of the papers accepted to the conference could actually be improved.

**Table 9.** demographic parity similarity and utility savings for our algorithms versus the baseline (Boolean).

| Method | Demographic Similarity | $Y_i$ | F-score |
|---|---|---|---|
| Overall Diversity | 89.15% | 93.47% | 91.26 |
| Multi-Faceted | 95.01% | 97.52% | 96.24 |
| Round Robin | 87.40% | 95.14% | 91.11 |

**Table 10.** demographic parity similarity and utility savings for our algorithms versus the baseline (Continuous).

| Method | Demographic Similarity | $Y_i$ | F-score |
|---|---|---|---|
| Overall Diversity | 94.80% | 102.49% | 98.27 |
| Multi-Faceted | 95.12% | 102.45% | 98.44 |
| Round Robin | 87.38% | 94.06% | 90.60 |

## 5.5 Discussion

In the previous sections, we discuss the evaluation for our approaches by comparing them to the baseline using Diversity Gain($D_G$), Utility Saving ($Y_i$), F-score metrics, and DemographicSimilarity. We compared three algorithms for our recommender system using Boolean and continuous weight features in the demographic profiles.

### 5.5.1 Diversity Gain Comparison

Table 11 summarizes the results for all experiment with each algorithm on both profile weights, evaluated using Diversity Gain, Utility loss, and the F-measure. From this we can see that the Overall Diversity method with Boolean profiles maximized diversity gain and the Overall Diversity method with continuous weights minimizes the utility loss. However, the Overall Diversity method with Boolean weights produces the highest F-score that balances these.

**Table 11.** Diversity Gain and Utility Saving with Boolean and Continuous weights profiles

|  | Profile | $D_G$ | $Y_i$ | F-score |
|---|---|---|---|---|
| Overall Diversity | Boolean | **64.58** | 93.47 | **76.39** |
|  | Continuous | 44.90 | **102.49** | 62.44 |
|  | Average | 54.74 | 97.98 | 69.41 |
| Multi-Faceted Diversity | Boolean | **46.00** | 97.52 | **62.51** |
|  | Continuous | 42.50 | **102.45** | 60.08 |
|  | Average | 44.25 | 99.98 | 61.29 |
| Round Robin | Boolean | **72.65** | **95.14** | **82.39** |
|  | Continuous | 66.80 | 94.06 | 78.12 |
|  | Average | 69.73 | 94.60 | 80.26 |
| Average | Boolean | 61.08 | 95.38 | 73.76 |
|  | Continuous | 51.40 | 99.67 | 66.88 |

**All the methods Comparison**

Averaged over both feature weights, the Round Robin measure produces higher diversity gain, 69.73% versus 44.25% and 54.74%. The Multi-Faceted Diversity method produces a smaller drop in utility on average, 0.02% versus 2.02%. However, when balancing diversity gain with utility drop, the Round Robin method still produces better results with an average F-score of 80.26 versus 69.41 and 61.29 for the Overall and Multi-Faceted approach.

**Boolean versus Continuous feature weights**

Averaged over the three methods, the Boolean profiles produces higher diversity gain, 61.08% versus 51.40% for Continuous. The Continuous profiles produces a little drop in utility on average, 0.33% versus 4.62% drop in utility for Boolean profiles. However, when balancing diversity gain with utility drop, the Boolean profiles still produces better results with an average F-score of 73.76 versus 66.88 for Continuous profiles.

### 5.5.2 Demographic Parity Comparison

Table 12 summarizes the results for all experiment with each algorithm on both profile weights, evaluated using Demographic Similarity, Utility loss, and the F-measure. From this we can see that the Multi-Faceted method with Continuous profiles maximized Demographic Similarity and the Overall Diversity method with continuous weights minimized the loss in utility. In fact, it actually produced a gain in utility! The Multi-Faceted method with Continuous weights also produces the highest F-score that balances these.

**Table 12.** Demographic Similarity with Boolean and Continuous weights profiles

|  | Profile | Demographic Similarity | $Y_i$ | F-score |
|---|---|---|---|---|
| Overall Diversity | Boolean | 89.15 | 93.47 | 91.26 |
|  | Continuous | **94.80** | **102.49** | **98.27** |
|  | Average | 91.97 | 97.98 | 94.76 |
| Multi-Faceted Diversity | Boolean | 95.01 | 97.52 | 96.24 |
|  | Continuous | **95.12** | **102.45** | **98.44** |
|  | Average | 95.06 | 99.98 | 97.34 |
| Round Robin | Boolean | **87.40** | **95.14** | **91.11** |
|  | Continuous | 87.38 | 94.06 | 90.60 |
|  | Average | 87.39 | 94.60 | 90.86 |
| Average | Boolean | 90.53 | 95.38 | 92.87 |
|  | Continuous | 92.43 | 99.67 | 95.77 |

**All the methods comparison**

Averaged over both feature weights, the Multi-Faceted Diversity measure produces higher Demographic Similarity, 95.06% versus 91.97% and 87.39% for the other two methods. The Multi-Faceted Diversity method produces a smaller drop in utility on average, 0.02% versus 2.02% and 5.4%. Also, when balancing diversity gain with utility drop, the Multi-Faceted Diversity method still produces better results with an average F-score of 97.34% versus 94.76% and 90.86% for the Overall Diversity approach and Round Robin.

**Boolean versus Continuous feature weights**

Averaged over both methods, the Continuous profiles produces higher Demographic Similarity, 92.43% versus 90.53% for Boolean profiles. The Continuous profiles produces a little drop in utility on average, 0.33% versus 4.62% drop in utility for Boolean profiles. However, when balancing diversity gain with utility drop, the Continuous profiles still produces better results with an average F-score of 95.77 versus 92.87 for Boolean profiles.

## 6 Conclusion

Our goal is maximizing diversity while reducing quality loss when recommending papers for inclusion in conferences. Thus, we provide new recommendation algorithms to select papers for publication in a conference. Our methods promote diversity by taking into account multiple demographic characteristics of authors in addition to paper quality. The majority of earlier research has concentrated on algorithms that ensure fairness based on a single Boolean attribute such as ethnicity, gender, or disability. In contrast, we profile the authors based on five attributes, i.e., gender, ethnicity, career stage, university rank, and geographic region. Furthermore, we extend previous work by comparing Boolean with Continuous-valued feature weights. To compare our algorithms for a real-world task, we created a dataset of authors whose papers were chosen for publishing at computer science conferences with varying impact factors to simulate papers judged by reviewers at various levels of acceptability in order to show our methodology.

We evaluated three algorithms: Overall Diversity, Multi-faceted Diversity, and Round Robin. The Overall Diversity method rates the papers based on an overall diversity score; the Multi-Faceted Diversity method chooses papers that satisfy the highest-priority demographic attribute participation goal first; and the Round Robin approach selects a single paper for each demographic attribute at a time. The resultant recommended papers were compared to the actual papers accepted in SIGCHI 2017 in terms of diversity gain and utility savings, where utility savings was measured based on how much, if at all, the h-index of the selected papers' authors declined. Our results showed that, based on the F-measure, the Round Robin approach with Boolean feature weights was the most effective strategy for maximizing diversity. It increased diversity by 72.6 percent with only a 4.8 percent reduction in utility. Also based on the F-measure, the Multifaceted Diversity algorithm with continuous features weights in the demographic profile was the most effective strategy for reaching demographic parity. With a 42.50% gain in diversity and a 2.45% boost in utility, it managed to achieve 95.12% similarity to the demographics of the pool of authors of all submitted papers. *This finding indicates that,*

*contrary to expectations, improving diversity actually raises the quality of publications that are accepted.*

In the future, we will investigate dynamic hill-climbing algorithms that modify the recommendation goals following each paper selection. Additionally, we will research the efficacy of fair deep learning approaches for paper recommendation.